\documentclass[11pt]{article}
\usepackage{moriond,epsfig}

     \newcommand{\pathnow}{}

\def\be{\begin{equation}}
\def\ee{\end{equation}}
\def\bea{\begin{eqnarray}}
\def\eea{\end{eqnarray}}

\begin{document}
  \hyphenation{strang-en-ess}
\vspace*{4cm}
\title{\uppercase{Strangeness and  QGP  freeze-out dynamics}}

\author{Johann Rafelski$^1$, Giorgio Torrieri$^1$ and Jean Letessier$^2$ }

\address{$^1$Department of Physics, University of Arizona, Tucson, AZ 85721\\
$^2$LPTHE, Universit\'e Paris 7,  2 Place Jussieu, F-75251 Cedex 05, France
}

\maketitle
\abstracts{
We compare chemical and thermal
analysis of the SPS Pb--Pb results at $158A$ GeV, 
and present a first chemical analysis of RHIC results. 
We show how  a combined 
analysis of several strange hadron resonances 
can be used  in a study of freeze-out dynamics.
}

\section{Introduction}
Strangeness signature of QGP originates in the observation that 
when color bonds are broken, the chemically (abundance) equilibrated 
deconfined state has an unusually high abundance of strange
quarks \cite{abundance}. 
Considering  the possibility that the  relatively small
size of the plasma fireball would suppress this enhancement,
It was shown that when the system size
is greater than about five elementary hadronic volumes \cite{RD80}
the physical properties of the hadronic system, including 
in particular strangeness enhancement, are nearly 
as expected for an infinite system. 
Kinetic study of the dynamical process 
of chemical strangeness equilibration  demonstrated that only the  
gluon component in the QGP is 
able to produce strangeness rapidly \cite{RM82}. 
Therefore strangeness enhancement is today considered to be 
related directly to  presence of gluons in QGP.

The high density of strangeness  in the reaction fireball 
favors formation of multi strange hadrons \cite{Raf82,RD83}, which 
are produced  rarely if only individual hadrons 
collide \cite{Koc85,KMR86}. In particular a large enhancement 
of multi strange antibaryons arising with a threshold behavior
as function of energy  has been proposed as  characteristic and 
nearly background-free signature of QGP \cite{Raf82}. 

A systematic strange antibaryon enhancement  has
in fact been observed, rising with strangeness content \cite{WA97p}.
Moreover there is now evidence that the enhancement of $\overline\Xi$
shows  a sudden onset when the number of participating (wounded)
nucleons exceeds 50 \cite{NA57Mor}. Similar results were reported for the
Kaon yields by the NA52 collaboration \cite{Kab99}. This threshold 
behavior arises for volumes which are relatively 
large and not a result of the smallness of the physical system 
(`canonical suppression' \cite{Red01}), thus 
arise from  opening up of novel reaction mechanisms, 
as is expected should QGP formation occur.

We also see in the experimental data that particles of very different 
properties are appearing with 
identical or similar $m_\bot$-spectra \cite{Ant00}.  
The  symmetry between strange
baryon and antibaryon spectra is strongly suggesting that 
the same reaction mechanism produces $\Lambda$ and $\overline\Lambda$ 
and $\Xi$ and $\overline\Xi$. This is understood readily if
 a dense fireball of matter formed in heavy ion
reactions  expands explosively, super cools, 
and in the end encounters a mechanical instability which facilitates 
sudden break up into hadrons  \cite{Raf00}.

Important in the understanding of the strange particle signatures of 
the QGP phase is the proper determination of the baseline of particle yield 
expected. The comparison of 
$AA$ (nucleus-nucleus) results should be always made with the 
$NA$ (nucleon-nucleus) collision system as the baseline, and in a wide range
the value of $A$ should not matter. Experimental results  demonstrate
that multi strange antibaryons are enhanced against this well defined
$NA$ baseline in a pattern expected in QGP hadronization. 
Yet it has been  argued  that strangeness signals of QGP are 
not unique \cite{Red01}, since  comparing $pp$  (proton-proton)
to $pA$ (proton-nucleus) interactions one observes a change of
production pattern of  strange particles.  Since 
in these reactions also a change in  non-strange  particle yields 
is observed  due to isospin selection rules
and shadowing, we believe that this line of taught is incorrect. 

We report in section \ref{freeze} that
thermal freeze-out occurs at the same condition found in
chemical analysis. We also consider the
present knowledge about strange hadrons at RHIC.
In section \ref{resonance}, we will 
introduce a method to evaluate  the lifespan of the hadronic
phase following formation of hadron multiplicity. The idea is to use
abundance of unstable resonances which have varying width and to 
determine  fraction  which becomes unobservable in
consideration of the re-scattering effects.


\section{Chemical and thermal freeze-out}\label{freeze}
\subsection{Strange hyperon $m_\bot$ spectra}\label{freezespectra}
In recent months experiment WA97 
determined the relative normalization of $m_\bot$-distribution for
strange particles $\Lambda,\,\overline\Lambda, \,\Xi,\,
\overline\Xi,\, \Omega+ \overline\Omega, \, K_s=(K^0+\overline{K^0})/2$
in four  centrality bins \cite{Ant00}.
We have since obtained a simultaneous description of the absolute yield 
(chemical freeze-out) and shape (thermal freeze-out) of these results \cite{Tor00}. 
Our strategy is to maximize the precision of the 
description of the final multi-particle hadron state employing
statistical methods. 
This requires that we introduce parameters which characterize
possible chemical non-equilibria, and  velocities  of matter
evolution. These latest results were obtained with
two velocities: a local  flow velocity $v$ of the fireball 
volume element where from particles emerge,
and hadronization surface (breakup) velocity which we refer to 
as $v_f^{\,-1}\equiv dt_f/dx_f$.

We have found, as is generally believed and expected, 
 that all hadron $m_\bot$-spectra are strongly influenced by 
resonance decays. We assume here that the resonance spectra are
not reequilibrated in rescattering. The final
particle distribution is composed of directly produced particles
and first generation decay products,
as no other contributing decays are known for hyperons, and 
hard kaons. Since the relative contributions of 
resonances and directly emitted particles are strongly 
temperature dependent, thermal analysis of hyperons 
converges to a well defined best temperature and velocities
of expansion and hadronization.

\begin{figure}[tb]
\centerline{
\psfig{width=8.cm,clip=1,figure=\pathnow 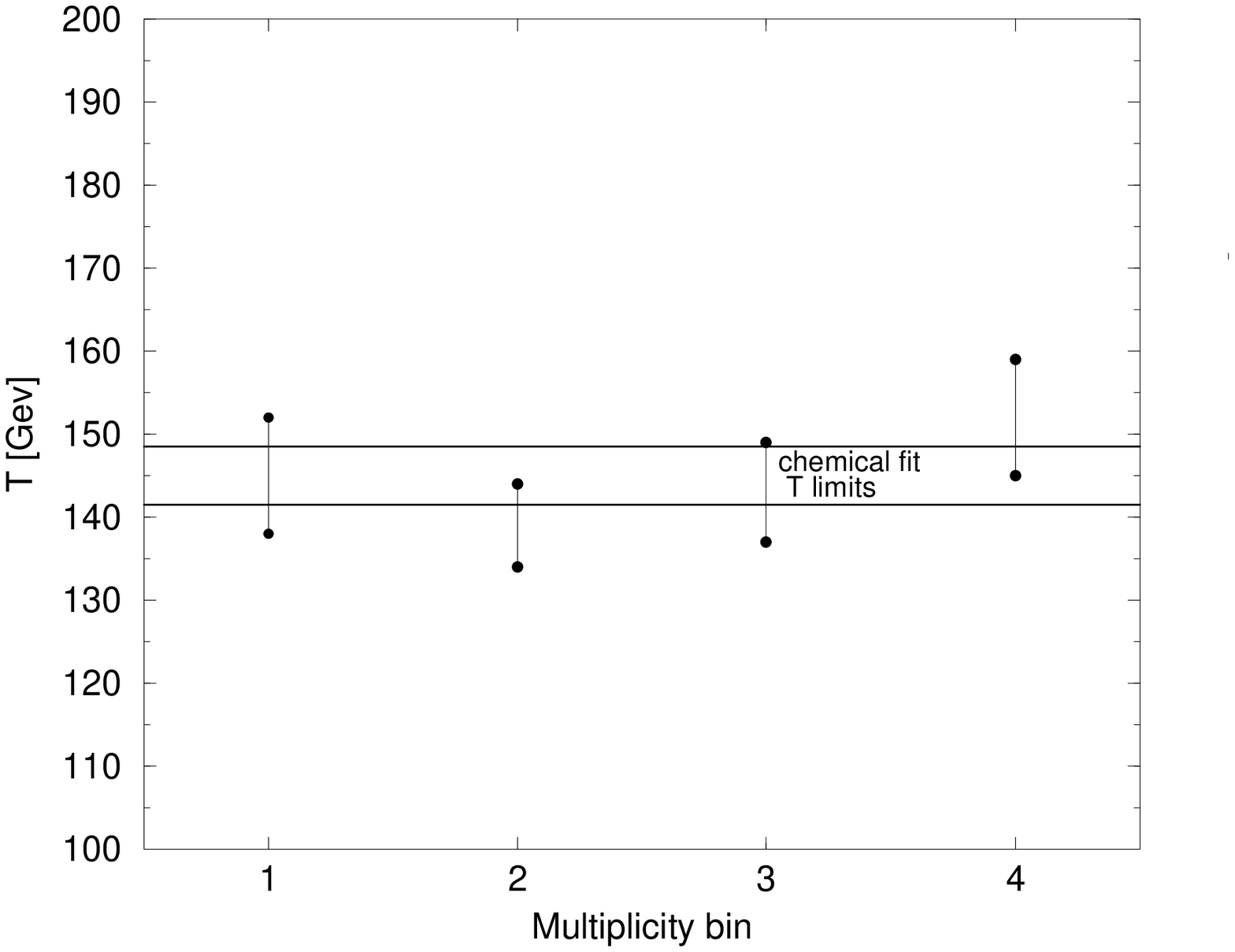}
\hspace*{-0.6cm}
\psfig{width=7.8cm,clip=1,figure=\pathnow 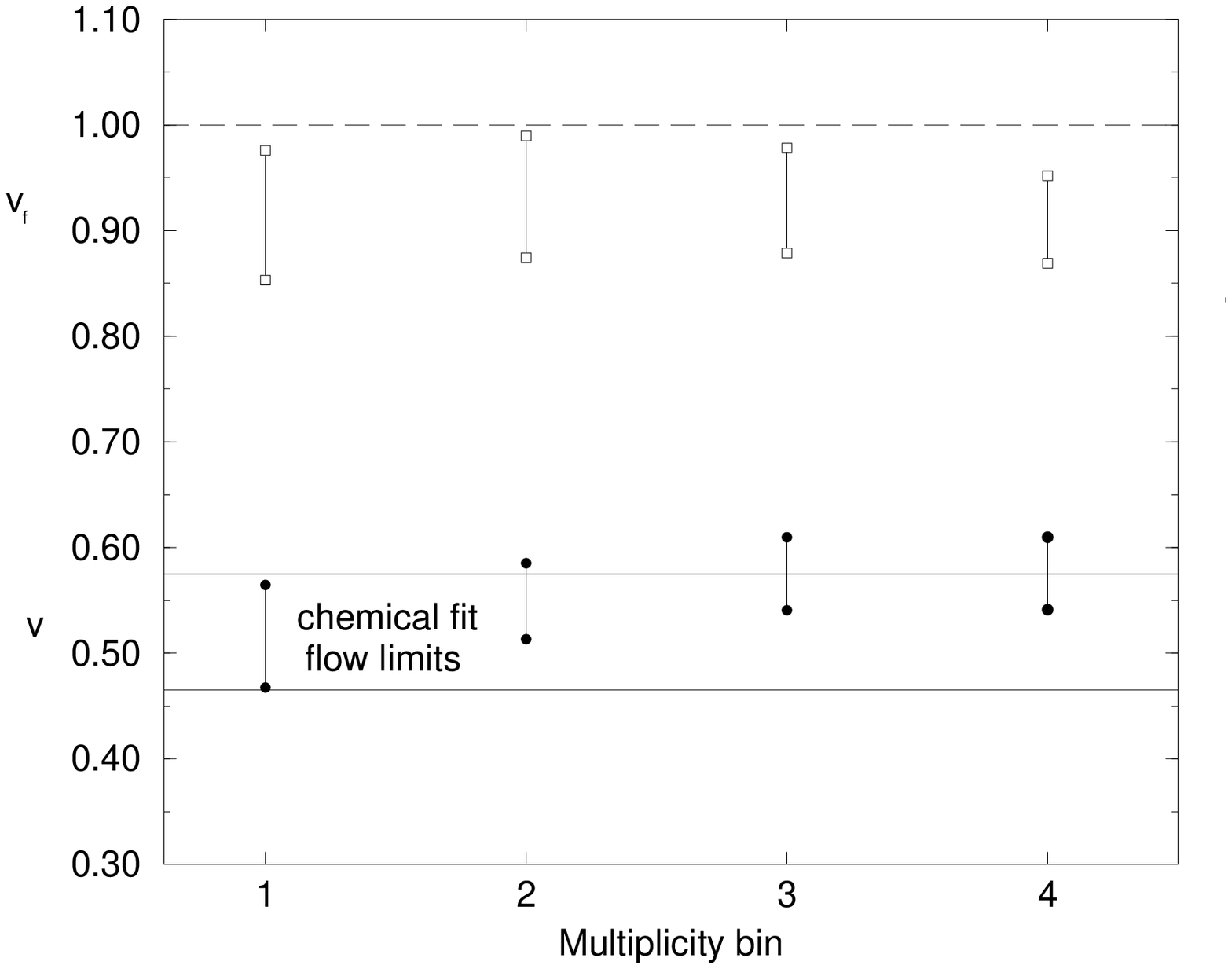}
}
\vspace*{-0.2cm}
\vspace*{0.2cm}
\caption{ 
Thermal freeze-out temperature $T$ (left) and 
flow velocity $v$ (bottom right) and 
break up (hadronization hyper-surface propagation) velocity $v_f$ 
(top right) for different collision centrality bins. 
Upper limit $v_f=1$ (dashed line) and 
chemical freeze-out analysis limits for $v$ (solid lines) are also shown.
\label{TdTTdv1v2}
}
\end{figure}

We present in Fig.\,\ref{TdTTdv1v2} these  freeze-out properties.
The solid horizontal lines 
delineate error range of  chemical freeze-out analysis 
we  present in  subsection \ref{freezechem}. 
We see that the thermal  freeze-out is
consistent with the purely chemical analysis of data
that included non-strange mesons and baryons.
The value of $v_f$ (top right) is near to 
velocity of light which is   consistent with the picture of a 
sudden breakup of the  fireball.

There is  no indication in Fig.\,\ref{TdTTdv1v2}
of a significant or systematic change of $T,v,v_f$ with centrality. 
This is consistent with the believe that the formation of the new state of 
matter at CERN is occurring in all centrality bins explored by the 
experiment WA97. It will be interesting to see if the low centrality 5th bin now
studied by experiment WA57 and which shows a different enhancement 
pattern  \cite{NA57Mor}, will also show different thermal freeze-out properties.

\begin{figure}[tb]
\vspace*{-0.7cm}
\centerline{\hskip 1.cm
\psfig{height=9.8cm,clip=,angle=-90,figure=\pathnow 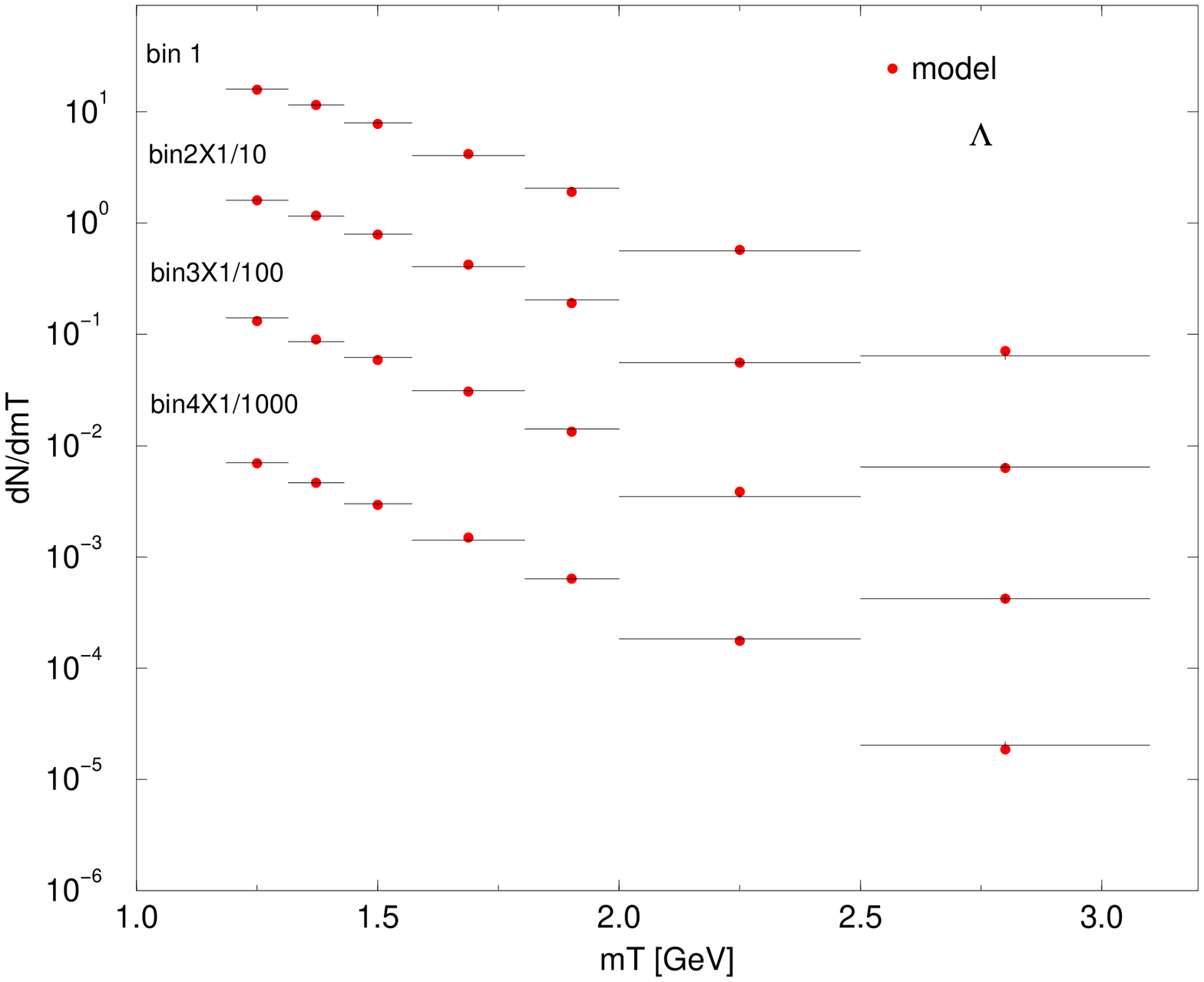}
\hspace*{-2.2cm}
\psfig{width=8.cm,clip=,angle=-90,figure=\pathnow 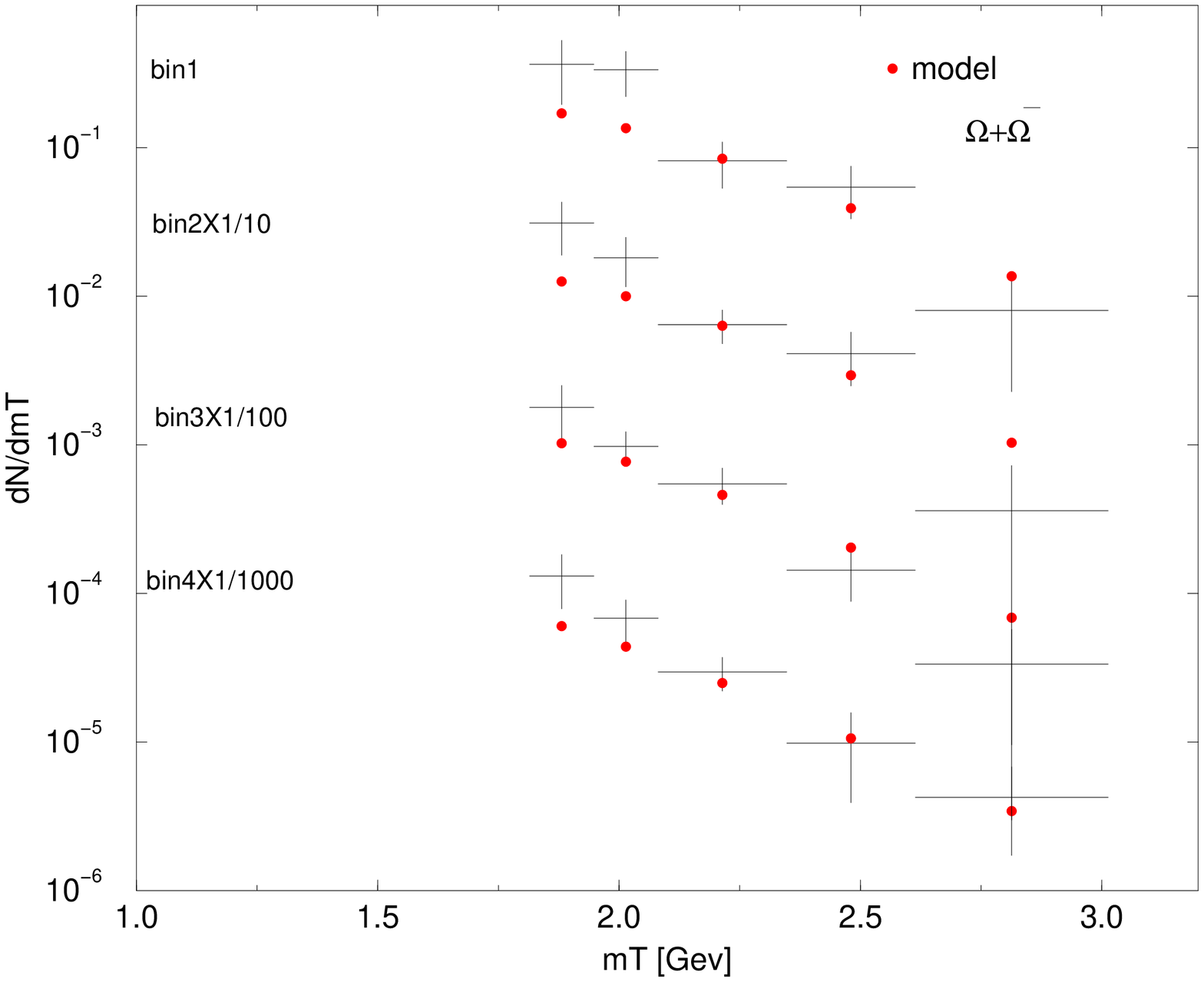}
}
\vspace*{-1.cm}
\caption{ 
Thermal analysis of $\Lambda$ (left) and $\Omega+\overline\Omega$ 
(right) $m_T$ spectra for different centrality of collision.
\label{TdLamtotOm}
}
\end{figure}

We show in Fig.\,\ref{TdLamtotOm} to left $\Lambda$-spectra which are
most precisely known, and to right $\Omega+\overline\Omega$-spectra which are
least precisely known. All other (anti)hyperon $m_\bot$-spectra 
($\overline\Lambda,\Xi,\overline\Lambda$) are described as well as 
we see it in the $\Lambda$-spectra.
We found that parameters found in  the analysis of hyperons and antihypoerons 
 predicted  correctly the  $K_s$ $m_\bot$-spectra.

Although in the purely chemical fit, we excluded the $\Omega,\overline\Omega$
yields due to their anomalous enhancement, we did include their spectra
in the thermal analysis.  In  all four centrality bins for the
 sum $\Omega+\overline\Omega$ we systematically  under predict the two lowest
$m_\bot$ data points, as is seen  in Fig.\,\ref{TdLamtotOm} right panel.
This low-$m_\bot$ excess also explains why the 
inverse $m_\bot$ slopes for $\Omega,\overline\Omega$ 
are reported to be smaller than the values seen in all other strange (anti)hyperons.
We note that the 1.5 s.d. deviations in the 
low  $m_\bot$-bins of the $\Omega+\overline\Omega$ 
spectrum translates into 3 s.d. 
deviations from the prediction of the statistical model chemical analysis. 

\begin{table}[tb]
\caption{\label{newfitqpbs}
Freeze-out conditions and physical properties of a hadronic 
matter fireball formed in Pb--Pb interactions at 
$\sqrt{s_{NN}}=17.2$ GeV, left 
column with, and right column without imposed strangeness balance. 
}
\vspace{0.4cm}
\begin{center}
\begin{tabular}{|l|c|c|}
\hline
                           &Pb$|_v^{\rm s,\gamma_q}$ &Pb$|_v^{\rm \gamma_q}$\\
$\chi^2_{\rm T};\ N;p;r$   &2.25;\ 10;\,3;\,2 &\hspace*{-0.5cm} 1.36;\ 10;\,4;\,2 \\
\hline
$T$ [MeV]             &150 $\pm$ 3     &145 $\pm$ 3.5 \\
$v$                     &0.57 $\pm$ 0.04  &0.52 $\pm$ 0.055\\
$\lambda_{q}$             &1.616 $\pm$ 0.025&1.625 $\pm$ 0.025\\
$\lambda_{s}$             &1.105$^*$         &1.095 $\pm$ 0.02\\
$\gamma_{q}$  & ${\gamma_{q}^c}^*\!=\!e^{m_\pi/2T_f}\!=\!1.61$  &${\gamma_{q}^c}^*\!=\!e^{m_\pi/2T_f}\!=\!1.59$\\
$\gamma_{s}/\gamma_{q}$   &1.02 $\pm$ 0.06  &1.02 $\pm$ 0.06\\
\hline 
$E_{f}^{in}/S_{f}$   &0.163 $\pm$ 0.01    &0.158 $\pm$ 0.01\\
${s}_{f}/b$               &0.68 $\pm$ 0.05  &0.69 $\pm$ 0.05\\
$({\bar s}_f-s_f)/b\ \ $  &0$^*$            &0.05 $\pm$ 0.05\\ 
\hline
\end{tabular}
\end{center}
\end{table}
\subsection{Global chemical freeze-out condition at SPS}\label{freezechem}
In our  chemical freeze-out analysis to which we compared the thermal
results in  Fig.\,\ref{TdTTdv1v2} there are a few theoretical refinements 
compared to earlier work \cite{Let00}, 
such as use of Fermi-Bose statistics throughout, more extensive 
resonance cascading. In the input data we omit the  
NA49 $\overline\Lambda/\bar p$ 
ratio and update the  NA49 $\phi$-yields. The total $\chi^2_{\rm T}$,
the number of measurements used $N$ the number of parameters being varied $p$ 
and the number of restrictions on data points $r$ are shown in 
heading of the  table \ref{newfitqpbs}. The values imply that our model 
has a  very high confidence level. 

In the upper section  of table \ref{newfitqpbs}, we show statistical 
model parameters which best describe the experimental results for
Pb--Pb data. We show in turn chemical freeze-out temperature,
$T$ [MeV], expansion velocity $v$, the light and strange quark fugacities
$\lambda_{q},\lambda_{s}$ and light quark phase space 
occupancy $\gamma_{q}$ and the ratio
strange to light quark ratio $\gamma_{s}/\gamma_{q}$. 
We fix $\gamma_{q}$ at the point of maximum pion entropy density 
${\gamma_{q}^c}=e^{m_\pi/2T_f}$ since this is the natural value to which the fit
converges once the Bose distribution for pions is used. 

It is interesting that in the Pb--Pb collisions  $\gamma_{s}/\gamma_{q}$
 is so close to unity, the often tacitly assumed value. 
In this detail the revised analysis differs more than 2 s.d. 
from our earlier results \cite{Let00}.
Only other notable difference is the prediction for
$\overline\Lambda/\bar p\simeq 0.5$ (not shown in table).

In the bottom section  of table \ref{newfitqpbs}, we show 
physical properties of the fireball derived from
the properties of the hadronic phase space:  $E_{f}^{in}/S_{f}$, the 
specific energy per entropy of the hadronizing volume element in local rest frame; 
${s}_{f}/b$  specific strangeness per baryon;
$({\bar s}_f-s_f)/b$ net strangeness
 of  the full hadron phase space characterized by these
statistical parameters. 

We see, in the bottom of the right column in table \ref{newfitqpbs}, 
that within error strangeness is balanced. 
In the first column of table \ref{newfitqpbs}
 we see that imposing exact strangeness balance increases
the chemical freeze-out temperature $T$ from 145 to 150 MeV. Insisting 
on exact balance may be 
an incorrect procedure since the WA97 central rapidity data, which are 
an important input into this analysis, are only known at central rapidity.
It is likely that the longitudinal flow of light quark content contributes to 
some mild $s$--$\bar s$-quark separation in rapidity. For this reason we 
normally consider the results presented in right 
column of table \ref{newfitqpbs} to be more representative of 
the freeze-out dynamics.

\subsection{First look at RHIC freeze-out}\label{freezeRHIC}
There is now first hadronic particle and 
strangeness data from RHIC $\sqrt{s_{NN}}=130$ GeV,
presented at QM2001 by the STAR collaboration \cite{STARkstar}. 
We draw the following conclusions from these results;
\begin{enumerate}
\item  from $\bar p/p=0.6\pm0.02=\lambda_q^{-6}$ 
it follows $\lambda_q=1.089$;
\item  and hence   $\mu_B=38$ MeV  (18\% of SPS value) at $T=150$ MeV.
If a hadronization at  $T=175$ MeV applies this value rises to $\mu_B=44$ MeV.
\item The ratios
$\overline\Lambda/\Lambda=0.73\pm 0.03=\lambda_s^{-2}\lambda_q^{-4}$ and
$\overline\Xi/\Xi=0.82\pm 0.08=\lambda_s^{-2}\lambda_q^{-4}$ are consistent
within 1.5\% with $\lambda_s=1$,  value expected for sudden hadronization.
\item $K^+/K^-=0.88\pm 0.06$  is also consistent within error with $\lambda_q=1.089$.
\item On the other hand
the ratio  $K^*/\overline{K^*}\simeq 1$ differs from $K/\overline{K}$
significantly. This suggests that $K^*,\overline{K^*} $ yields are
influenced at the level of 10\% 
by `in hadronization' decay product re-scattering 
in an asymmetric way. 
\item Thus
$K^*, \overline{K^*}$ should not be used to fix $T$ 
using the ratios $ K^* /h^-$ and  $ \overline{K^*} /h^-$.
\item The ratio  $ \bar p/\pi=8\%$ cannot be used to fix $T$ 
since $ \bar p$ yield contains undetermined hyperon feed \cite{Raf99}.
\item  The ratio $K^-/\pi^-$ does not suffice to fix the
temperature: we need at least 3 reliable yield ratios as we must also
fix $\gamma_q,\gamma_s$:  $K^-/\pi^-=15\%=f(T)\gamma_s/\gamma_d$.
\end{enumerate}
We conclude that the first RHIC results allow to understand the magnitude of
chemical potentials $\mu_s=0, \mu_b=38$ MeV, 
but $T$ and $\gamma_q,\gamma_s$ cannot yet be fixed.
Given the rescattering phenomena of resonances one cannot do a global
analysis without stable strange hadron yields, akin to the situation
we have at SPS energy range. Thus the final analysis must await the time 
these results become available.  On the other hand the strong 
presence of observable resonances in hadronic
final state as reported by the STAR experiment 
implies that hadronization has occurred in a sudden fashion, as has 
been seen at SPS. Other  RHIC results such as correlation analysis, are
also strongly suggestive of sudden break-up/hadronization.

The major departure at RHIC from
SPS physics is the great strangeness density. 
We note that:
$$
\frac{dN_{K^+}}{dy}\vert_{y=0}=35\pm3.5\,,
\quad 
\frac{dN_{K^-}}{dy}\vert_{y=0}=30\pm3\,.
$$
Total strangeness $(\bar s)$ 
yield depends on unmeasured hyperons. Model calculations 
suggest more than 20\%. Hence:
$$
\frac{d\bar s}{dy}\vert_{y=0} > 85\pm9\,,\quad \mbox{compare\ to\ }\quad
\frac{d\pi^+}{dy}\simeq  \frac{d\pi^-}{dy} \simeq 235.
$$ 
Under these conditions calculations suggest that 
$\bar s/b\simeq 8$ (11--12 times greater than at 
$\sqrt{s_{NN}}= 17.2 \,A$ GeV SPS Pb--Pb). 

Given this immense strangeness rapidity yield it is very 
difficult to imagine that among three quarks which coalesce to make
a baryon there is  no strange quark! Hence we predict that most
baryons and antibaryons produced will carry strangeness \cite{Raf99}. 
Thus non-strange nucleons and antinucleons are strongly contaminated 
by hyperon decay feed, and at this time the reported nucleon RHIC results
cannot be used in order to characterize freeze-out conditions.

\section{Resonances and freeze-out dynamics}\label{resonance}
We consider strange hadron resonance production as probe
freeze-out dynamics\cite{Tor01,Raf01}.
The $\Lambda(1520)$ abundance yield is found about 2 times smaller
than expectations based on the yield 
extrapolated from nucleon-nucleon reactions \cite{Mar01}. This is 
to be compared with the  enhancement by factor 2.5 of 
$\Lambda$-production.  A possible explanation for
this effective suppression by a factor 5 is that the
decay products ($\pi,\Lambda$)  have re-scattered and thus their momenta
did not allow  to reconstruct this state in an invariant mass analysis.
In a study of the 
rescattering of the resonance products we found that if the
resonance decay occurs in fireball matter, one of the decay products 
will in general rescatter and thus the resonance will not be observable 
in the reconstruction of the invariant mass \cite{Tor01}. 

A back of envelope calculation based on exponential population attenuation
suggests that if the observable yield of  $\Lambda(1520)$ is reduced 
by factor 5, the observable yield of $K^*(892)$ with much greater 
width, $\Gamma_{K^*(892)}=50$ MeV, should be suppressed by a factor 15.
However, both SPS \cite{NA49Res2} and RHIC experiments \cite{STARkstar}
report measurement of  $K^*(892)$ signal. 
A possible explanation is that in matter the 
lifespan of $\Lambda(1520)$ can be quenched in collisions such as
$ \pi+\Lambda(1520)\to \Sigma^*\to   \pi+\Lambda$\,. This is possible
since $\Gamma_{\Lambda(1520)}$  is  small due to
need for angular $L=2$ partial wave in its decay.

\begin{figure}[tb]
\begin{center}
\centerline{\hspace*{.1cm}
\psfig{width=7.5cm,clip=1,figure=\pathnow 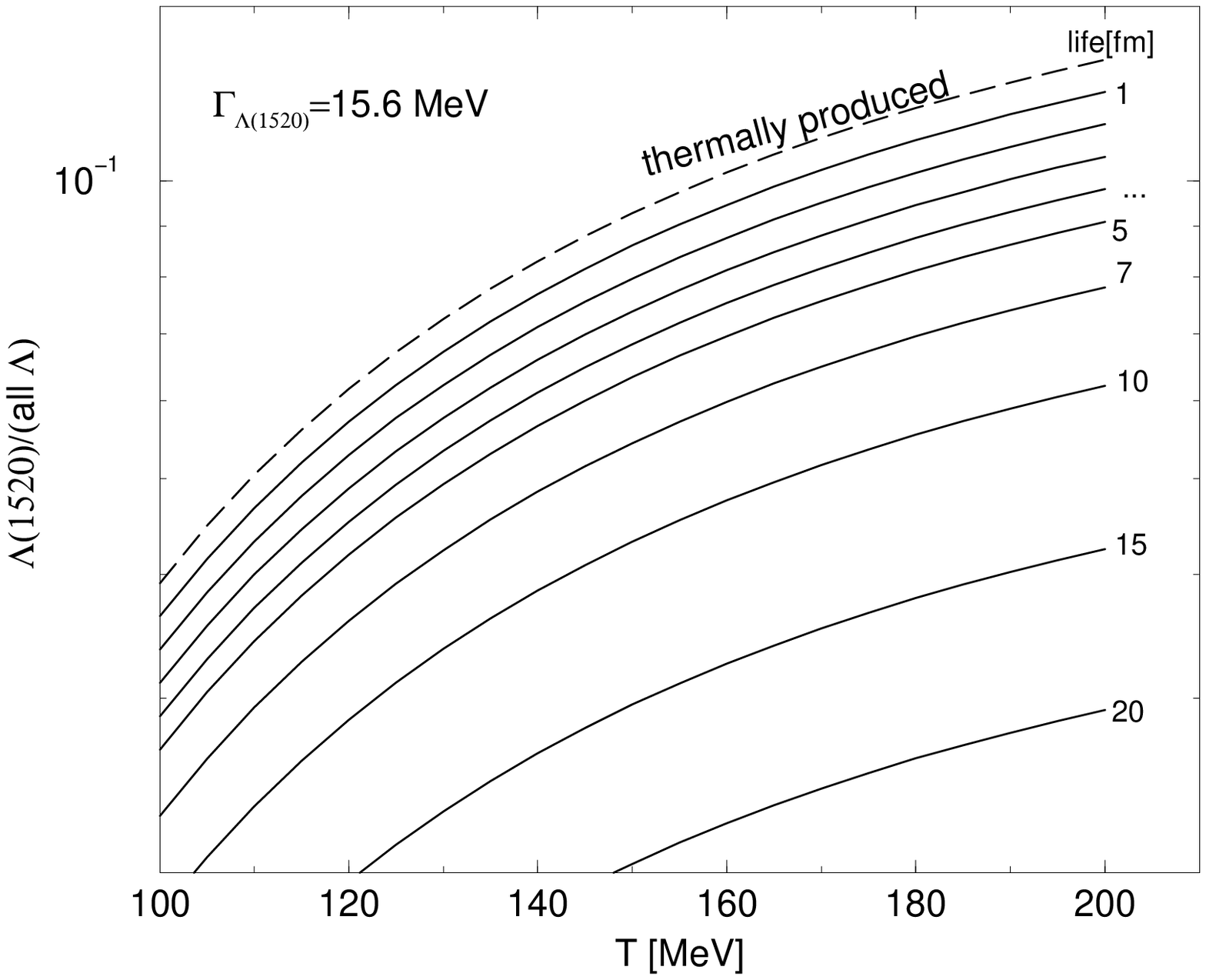}
\hspace*{0.1cm}
\psfig{width=7.5cm,clip=1,figure=\pathnow 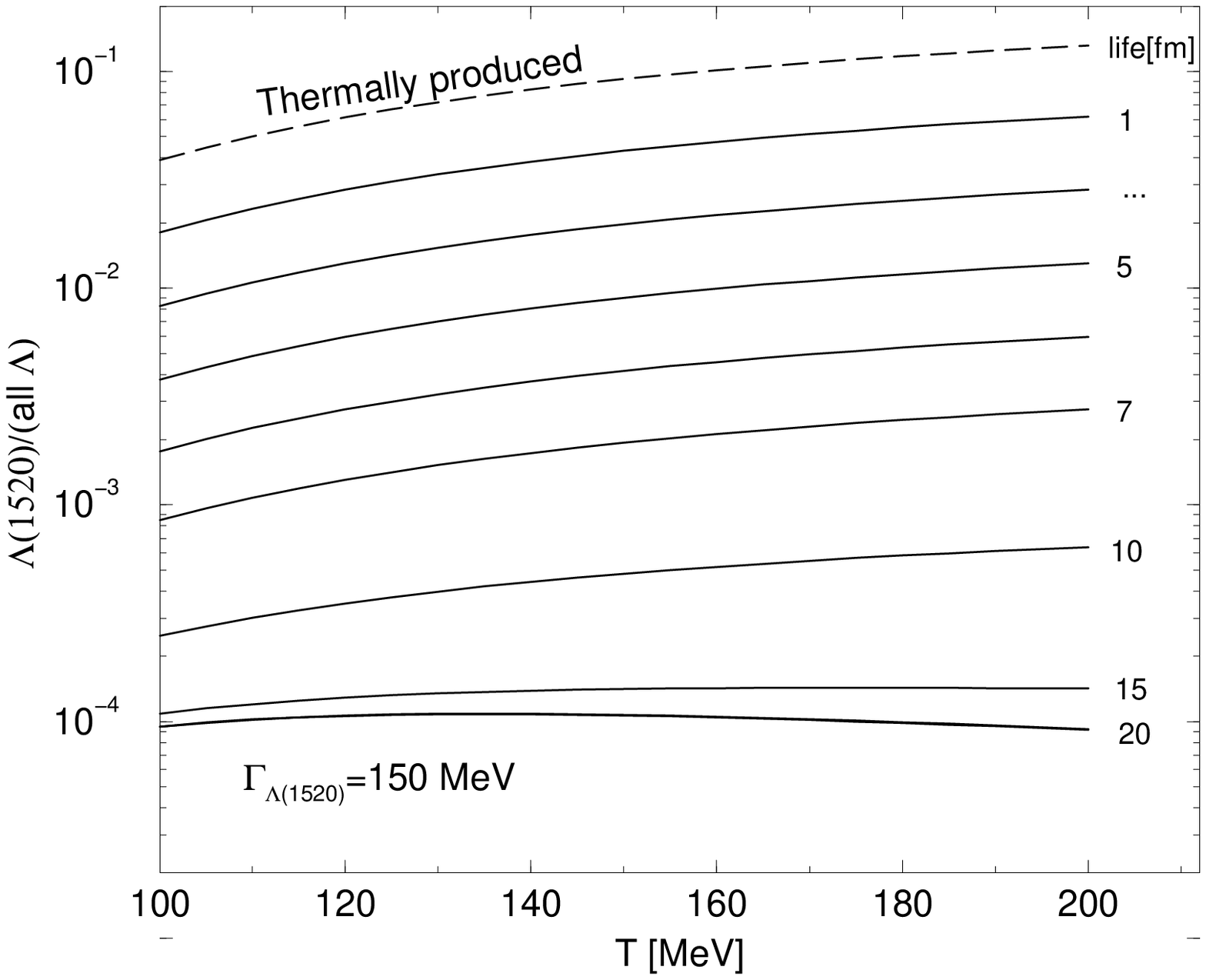}
}
\vspace*{-0.21cm}
\caption{Relative $\Lambda(1520)/(\mbox{all }\Lambda)$ yield as function
of freeze-out temperature $T$. Dashed - thermal yield, solid lines: 
observable yield for evolution  lasting the time shown (1....20 fm)
in an opaque medium. Left: natural resonance width 
$\Gamma_{\Lambda(1520)}=15.6$ MeV, right: 
quenched $\Gamma_{\Lambda(1520)}=150$ MeV.  
 \label{LamRes}}
\end{center}
\end{figure}

We show  in Fig.\,\ref{LamRes} how quenching impacts 
$\Lambda(1520)/(\mbox{all }\Lambda)$ yield:  left
panel is for the natural width $\Gamma_{\Lambda(1520)}=15.6$ MeV,
right panel is for a width quenched to 
$150$ MeV. NA49 has just reported
$\Lambda(1520)/(\mbox{all }\Lambda)=0.025\pm 0.008$ 
which is barely if at all compatible with the unquenched result, since it implies
an extremely long hadronization time of about $20\pm5$ fm/c (depending on
freeze-out temperature) which is incompatible with other 
experimental results. On the other hand, 
after introduction of a quenched 
resonance width  the experimental result 
is compatible for all freeze-out temperatures with 
a sudden hadronization model.

\begin{figure}[tbp]
\begin{center}
\vspace*{-2.cm}
\centerline{\hspace*{1.4cm}
\psfig{width=9.2cm,clip=1,figure=\pathnow 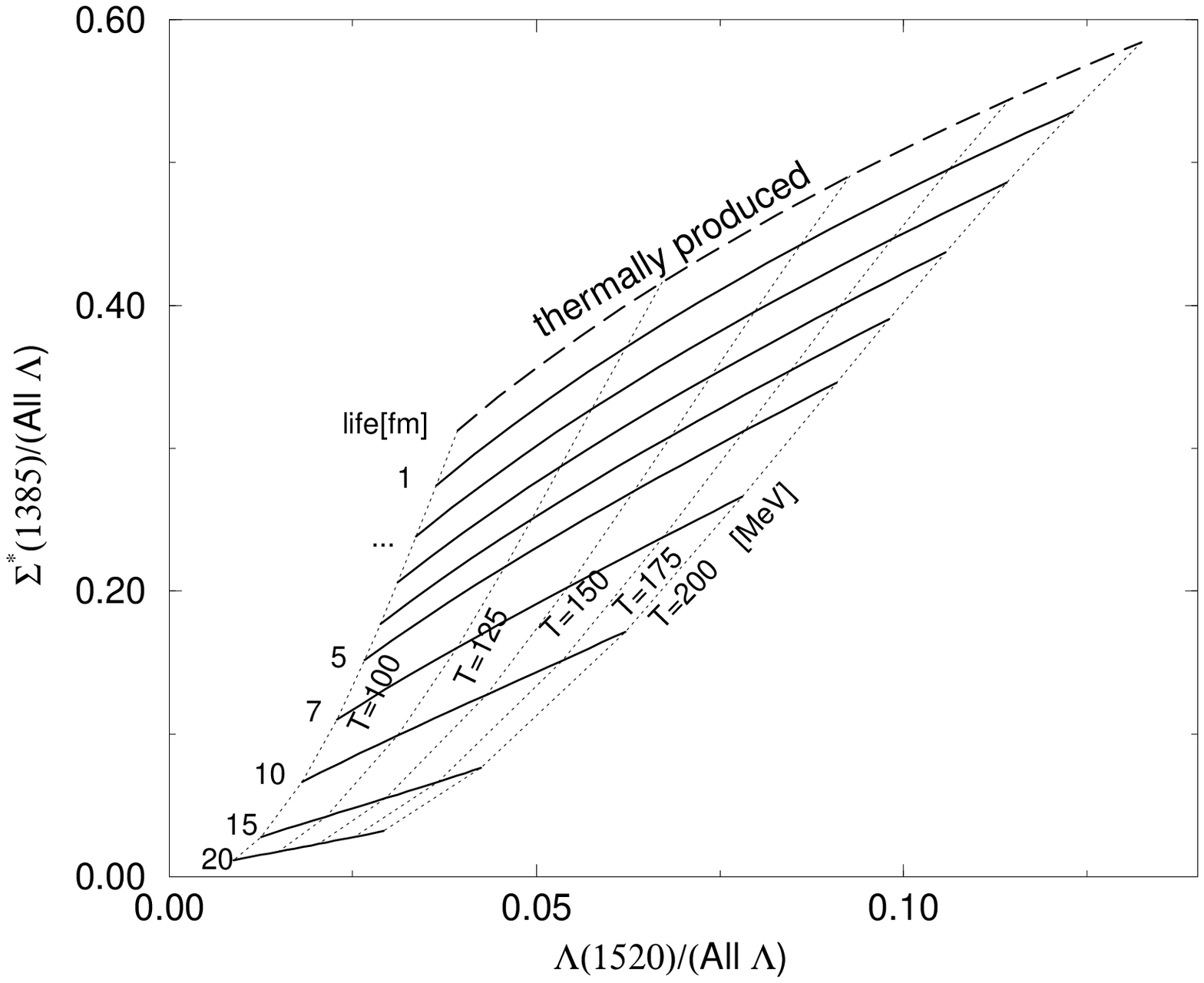 }
\hspace*{-1.5cm}
\psfig{width=9.2cm,clip=1,figure=\pathnow 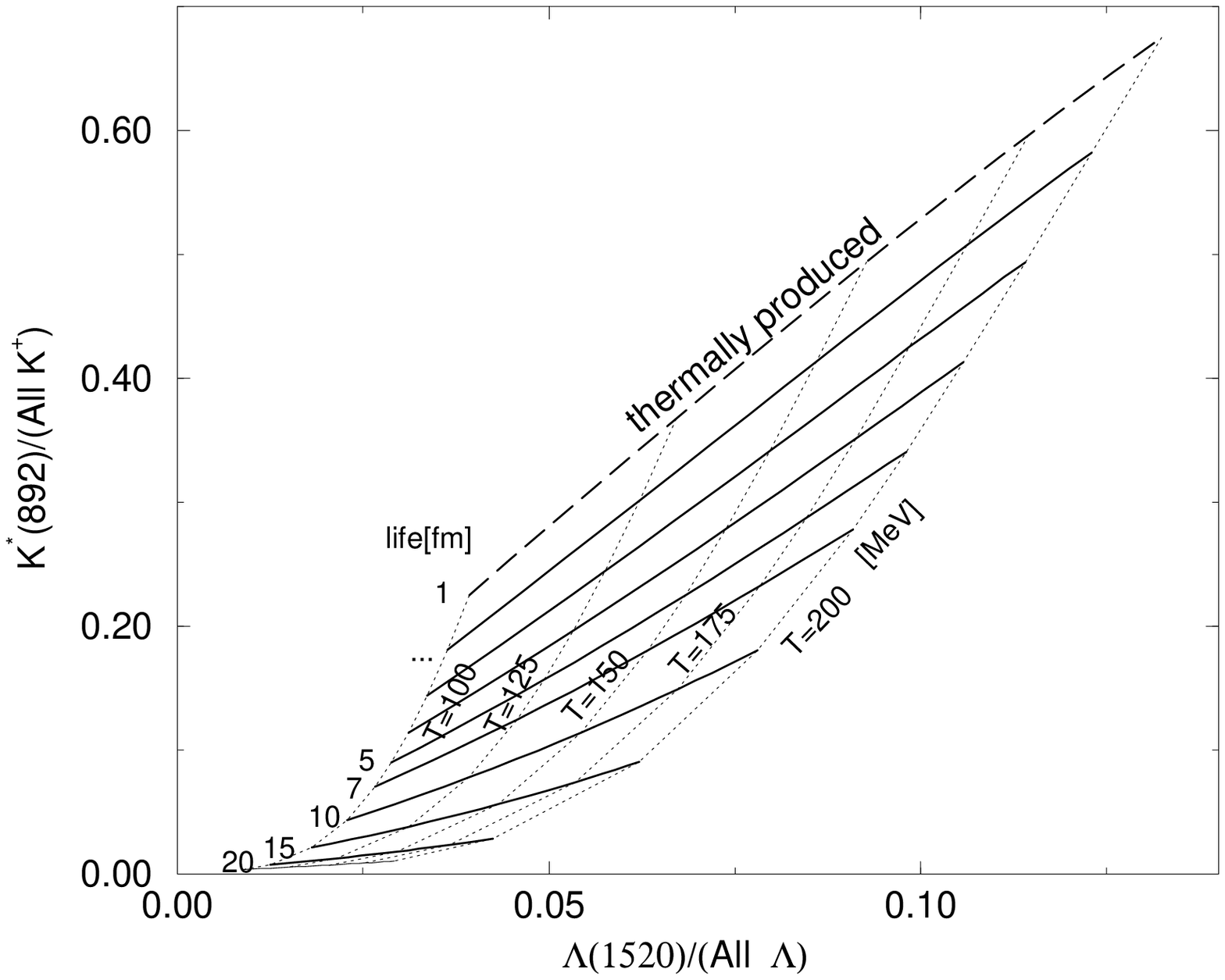}
}
\caption{Dependence of the combined $\Sigma^{*}$/(all $\Lambda$)with
$\Lambda(1520)$/(all $\Lambda$)  (left) and  
$K^*(892)/$(all K)  with $\Lambda(1520)$/(all $\Lambda$) (right)
resonance production  on the chemical freeze-out temperature
and hadron matter lifetime.}
\label{projdiagall}
\end{center}
\end{figure}

The study  of several strange hadron resonances  {\it e.g.}
$\Lambda(1520), K^*(892), \Sigma^*(1385)$  
$\Gamma_{\Sigma^*(1385)}=35$ MeV, provides a tool capable
of probing conditions at particle freeze-out. 
We recall that $\Sigma^*(1385)$ decays 
into $\Lambda$ and is  expected to be produced more abundantly 
than $\Lambda(1520)$ in a hadronic fireball due to it's high 
degeneracy factor and smaller mass. 
How a systematic approach will  work is shown in  Fig.~\ref{projdiagall}
which shows relative yield of one resonance as function of another, here 
presented for their natural widths.
As indicated from top to bottom in the grid, the 
lifespan in  matter increases, while from
left to right the temperature of chemical freeze-out increases. 

\section{Summary and conclusions}
To close we look at a few highlights of our report. 
Our thermal freeze-out analysis confirms that
CERN-SPS  results decisively show  interesting and new 
physics, and confirms the reaction picture of
a suddenly hadronizing QGP-fireball with both
chemical and thermal freeze-out being the same. 
Thermal freeze-out 
condition for strange hadrons (K$^0_s, \Lambda, \overline\Lambda,
\Xi, \overline\Xi$)  agree within error with chemical freeze-out
and we have confirmed the  freeze-out temperature $T\simeq 145$\,MeV.

We  were able to determine the freeze-out surface
dynamics and have shown that the break-up velocity $v_f$ is nearly
velocity of light, as would be expected in  a sudden breakup of a
QGP fireball. We found  that the experimental production data 
of $\Omega+\overline\Omega$ 
has a noticeable systematic low-$p_\bot$ enhancement anomaly 
present in all centrality bins. This result shows that it is not a
different temperature of freeze-out  of $\Omega+\overline\Omega$ 
 that leads to more enhanced yield, but a soft momentum secondary source
which contributes almost equal number of soft $\Omega+\overline\Omega$
compared to the systematic yield predicted by the other strange hadrons.

We have presented in section \ref{resonance} results 
on strange hadron resonance production and argued that a study of
several resonances with considerably different physical
properties must be  used in a study of freeze-out dynamics of QGP.
Strange resonances are easier to explore, since their decay 
involve rarer strange hadrons and thus the backgrounds are smaller. 
Moreover, the detectability of the naturally wide non-strange
resonances is always relatively small, except if (very) sudden hadronization
applies. For this reason it will be quite interesting
to see if  $\Delta(1230)$ can be observed at all, as this
would be only possible if  chemical and thermal freeze-out 
conditions are truly coincident. 

This discussion of how resonances help to 
understand the hadronization dynamics is a beginning of a complex
analysis which will occur in interaction with experimental 
results. We saw  that observable strange resonance yields can
vary widely depending on conditions which should allow a detailed study of 
QGP freeze-out dynamics. We believe considering $\Lambda(1520)$ result 
that in-matter resonance lifetime quenching is significant. 

\section*{Acknowledgments}
Work supported in part by a grant from the U.S. Department of
Energy,  DE-FG03-95ER40937. Laboratoire de Physique Th\'eorique 
et Hautes Energies, University Paris 6 and 7, is supported 
by CNRS as Unit\'e Mixte de Recherche, UMR7589.


\section*{References}
\begin{thebibliography}{99}
\bibitem{abundance}
 J. Rafelski, 
 pp.\,282--324, in  {\it Future Relativistic Heavy Ion  Experiments}, 
R. Bock and R. Stock, Eds., GSI Report 1981-6; 
J. Rafelski, and R. Hagedorn, 
pp.\,253--272 in: {\it Statistical Mechanics of Quarks and Hadrons},
H. Satz, ed.; (North Holland, Amsterdam, 1981);
J. Rafelski, 
{\it Nucl. Physics} A {\bf 374}, 489c (1982).

\bibitem{RD80}
J. Rafelski and M. Danos,  
{\it Phys.  Lett.} B {\bf 97}, 279 (1980). 

\bibitem{RM82}
{J. Rafelski and B. M\"uller}, 
{\it Phys. Rev. Lett} {\bf 48}, 1066 (1982); {\bf 56}, 2334E (1986);\\
{P. Koch, B. M\"uller and J. Rafelski},
{\it Z. Phys.} A {\bf 324}, 453 (1986).

\bibitem{Raf82}
J. Rafelski, 
{\it Phys. Rep.} {\bf 88}, 331 (1982).

\bibitem{RD83}
J. Rafelski and M. Danos, 
{\it Perspectives in High Energy Nuclear Collisions}, 
NBS-IR 83-2725 Monograph, U.S. Department of Commerce, 
National Bureau of Standards, June 1983;\\
Updated version  appeared  in 
{\it Nuclear Matter under Extreme Conditions}, D. Heiss, Ed.,
Springer Lecture Notes in Physics {\bf 231}, pp.\,362-455 (1985).

\bibitem{Koc85}
P.  Koch and J. Rafelski, {\it Nucl. Phys.} A {\bf 444}, 678 (1985).

\bibitem{KMR86}
P.~Koch, B.~M\"uller and J.~Rafelski, 
{\it Phys. Rep.} {\bf 142}, 167 (1986).



\bibitem{WA97p}
F. Antinori {\it et al.},  WA97 Collaboration 
{\it Nucl. Phys.} A {\bf 663}, 717 (2000);\\
E. Andersen {\it et al.}, WA97   collaboration,
{\it Phys. Lett.} B {\bf 433}, 209 (1998);\\
E. Andersen {\it et al.}, WA97  collaboration, 
{\it Phys. Lett.} B  {\bf 449}, 401 (1999).

\bibitem{NA57Mor}
D. Elia, for NA57 collaboration, 
presentation at {\it Rencontres de Moriond} (in this volume).

\bibitem{Kab99}
S. Kabana  {\it et al.}, NA52 collaboration,
{\it Nucl. Phys.} A {\bf 661}, 370c (1999);\\
S. Kabana  {\it et al.}, NA52 collaboration,
{\it J. Phys.} G {\it Nucl. Part. Phys.} {\bf 25}, 217 (1999).

\bibitem{Red01}
K. Redlich, presentation at {\it Rencontres de Moriond} (in this volume).

\bibitem{Ant00}
F. Antinori  {\it et al.},  WA97 Collaboration 
{\it Eur. Phys. J.} C {\bf 14}, 633, (2000), and
private communication.


\bibitem{Raf00}
J. Rafelski and J. Letessier, 
{\it Phys. Rev. Lett.}  {\bf 85}, 4695 (2000).

\bibitem{Tor00}
G. Torrieri, and J. Rafelski, 
\textit{Search for QGP and thermal freezeout of strange hadrons}
 hep-ph/0012102, submitted to {\it New J. of Phys.}.

\bibitem{Let00}
J. Letessier and J. Rafelski,
{\it Int. J. Mod. Phys.} E {\bf 9}, 107, (2000), and
references therein.


\bibitem{Raf99}
J. Rafelski and J. Letessier, {\it Phys. Lett} B {\bf 409}, 12 (1999).

\bibitem{STARkstar}
Zhangbu Xu, for the STAR Collaboration, 
``Resonance Studies at STAR''
nucl-ex/0104001, to appear in {\it  Nucl. Phys.} A (2001); and\\
Plenary session QM2001 presentation: 
Helen Caines for the STAR collaboration

\bibitem{Tor01}
G. Torrieri, and J. Rafelski, 
``{\it Strange hadron resonances as a signature of freeze-out dynamics}''
hep-ph/0103149, {\it Phys. Lett.} B. in press.

\bibitem{Raf01}
J. Rafelski, J. Letessier and G. Torrieri,
``{\it Strange hadrons and theri resonances: a diagnostic
tool of QGP freeze-out dynamics}''
nucl-ph/0104042. 


\bibitem{Mar01}
Ch. Markert, PhD thesis,  available at\\
na49info.cern.ch/cgi-bin/wwwd-util/NA49/NOTE?257\ .

\bibitem{NA49Res2}
V. Friese for the NA49 collaboration, QM2001 presentation.


\end{thebibliography}
\end{document}